\documentstyle{mn2e}
\input psfig

\def\solmas{{M$_\odot$}}
\def\simless{\mathbin{\lower 3pt\hbox
   {$\rlap{\raise 5pt\hbox{$\char'074$}}\mathchar"7218$}}}   % < or of order
\def\simgreat{\mathbin{\lower 3pt\hbox
   {$\rlap{\raise 5pt\hbox{$\char'076$}}\mathchar"7218$}}}   % > or of order
\def\etal{{\rm et al.}}

\def\solmas{{M$_\odot$}}
\def\solm{{M_\odot}}

\def\apj{{ApJ}}

\def\mnras{{MNRAS}}

\def\ARAA{{ARA\&A}}

  \makeatletter
  \ifx\CUP@mtlplain@loaded\undefined  % this is a font switch for CUP
  \makeatother  
    
  \else
    
  \fi

\loadboldmathitalic % loads bold math italic
\loadboldgreek      % loads bold greek
  
\makeatletter
\ifx\CUP@mtlplain@loaded\undefined  % this is a font switch for CUP
  \newfont\bit{cmbxti10 at 9pt}
\else
  \newfont\bit{mtbxti10 at 9pt}
\fi
\makeatother

\def\LaTeX{L\kern-.36em\raise.3ex\hbox{a}\kern-.15em
    T\kern-.1667em\lower.7ex\hbox{E}\kern-.125emX}

\newcommand{\gsim}{\mathrel{\hbox{\rlap{\lower.55ex \hbox {$\sim$}}
                   \kern-.3em \raise.4ex \hbox{$>$}}}}
\newcommand{\lsim}{\mathrel{\hbox{\rlap{\lower.55ex \hbox {$\sim$}}
                   \kern-.3em \raise.4ex \hbox{$<$}}}}

\title[Generation of velocity dispersion in GMCs] {Spiral shocks, triggering of star formation and the velocity dispersion in Giant Molecular Clouds}
\author[I. A. Bonnell \etal]
  {I. A. Bonnell$^1$\thanks{E-mail: iab1@st-and.ac.uk},  C.L. Dobbs$^1$, T. P. Robitaille$^1$  \& J.E. Pringle$^2$ \\
$^1$ School of Physics and
  Astronomy, University of St Andrews, North Haugh, St Andrews, Fife,
  KY16 9SS. \\
$^2$ Institute of Astronomy, Madingley Road, Cambridge, CB3 0HA \\ }

\date{\today}

\begin{document}

\maketitle

\begin{abstract}

We present numerical simulations of the passage of gas through a
galactic spiral shock and the subsequent formation of giant molecular
clouds (GMCs), and the triggering of star formation.  In these
simulations, we take account of the observed inhomogeneity, or
clumpiness, of the pre-shock interstellar medium. As might be
expected, the spiral shock forms dense clouds while dissipating
kinetic energy, producing regions that are locally gravitationally
bound and collapse to form stars. But the effect of the clumpiness of
gas as it passes through the shock is to generate chaotic internal
motions in the gas. The kinematics of these motions are found to agree
with the observed velocity-dispersion/size relation found in
star-forming regions. 
 In contrast to the standard picture where
continuously driven turbulence generates the density inhomogeneities
in star-forming clouds, we find here that it is the clumpiness of the
interstellar gas that produces the chaotic motions as it passes
through the spiral shock and initiates the star formation process. 
The velocity dispersion can be understood as
being due to the random mass loading of clumps as they converge
in the spiral shock. 
Within these clouds both the timescale for the decay of these motions,
and the timescale for forming stars, are comparable to the clouds'
dynamical lifetimes. In this model there is no need for any internal or external continuous driving 
mechanism for the 'turbulence'.  In addition,
the coupling of the clouds' internal kinematics to their externally
triggered formation removes the need for the clouds to be
self-gravitating. Indeed, while clearly some parts of the clouds are
self-gravitating and able to form stars, most of the molecular
material remains gravitationally unbound. This can provide a simple
explanation for the low efficiency of star formation.

\end{abstract}

\begin{keywords}
stars: formation --- galaxies: ISM --- ISM: clouds --- galaxies: star clusters --- galaxies: kinematics
and dynamics --- open              clusters and associations: general.
\end{keywords}

\section{Introduction}

Our understanding of the star formation process has recently undergone
a paradigm shift from the earlier quasi-static picture where the
environment plays no role, to an awareness that star formation is an
extremely dynamical process where the local environment, and
interactions, play a dominant role in determining the resultant
stellar properties (Hartmann \etal~2001; Larson~2003; Mac Low \&
Klessen~2004; Bonnell \etal~2001).  Traditionally, giant molecular
clouds (GMCs), the sites of star formation, were thought to be
long-lived entities (ages $>10$ dynamical times) in order to explain
the low star formation rate in our Galaxy (Zuckerman \& Evans~1974;
Blitz \& Shu~1980; Leisawitz~1990).  Recently, it has become clear
that observational evidence from young stellar populations implies a
relatively fast star formation process whereby clouds appear, form
stars and then disperse on their local dynamical times of a few
million years (Elmegreen~2000; Hartmann \etal~2001).  One of the basic
properties of these clouds is that they display internal chaotic
motions which are highly supersonic. These motions are usually
interpreted as 'turbulence', and observations suggest that any driving
of the observed turbulence should come from the largest scales of the
GMC (Brunt~2003).  A succession of numerical simulations (Mac-Low et
al., 1998; Padoan \& Nordlund, 1999; Ostriker, Gammie \& Stone, 1999)
have demonstrated that such supersonic motions, in the absence of some
continuous driving mechanism, dissipate on one or two dynamical
timescales.  The realisation that the lifetimes of these clouds are
comparatively short then suggests that the formation mechanism for
GMCs is a dynamical one (Roberts~1969; Shu \etal~1972; Pringle \etal~2001). One implication of such a mechanism is that gravitational
clouds need not control their own evolution in the sense that,
contrary to the usual assumption, they need not be self-gravitating
entities (although of course some parts of then do need to be in order
to form stars), nor in virial equilibrium. Thus the internal motions
which we see may be kinematic, rather than dynamic (Pringle
\etal~2001).  If molecular clouds are globally gravitationally
unbound, the perceived problems of their low star formation
efficiencies and supposed long lifetimes are erased
(Elmegreen~2000). Recent simulations of unbound molecular clouds have
shown how star formation can proceed while the majority of the gas
escapes due to its excess kinetic energy (Clark \& Bonnell~ 2004;
Clark \etal~2005).

Star formation has long been known to occur primarily in the spiral
arms of disc galaxies (Baade 1963). Spiral arms are denoted by the
presence of young stars, HII regions, dust and giant molecular clouds,
all signatures of the star formation process (van den Bergh~1964;
Schweizer~1976; Bash, Green \& Peters~1977; Elmegreen \&
Elmegreen~1983; Rumstay \& Kaufman~1983; Ferguson \etal~1998).  What
is still unclear is the exact role of the spiral arms in inducing the
star formation.  Is it simply that the higher surface density due to
the orbit crossing is sufficient to initiate star formation, as in a
Schmidt law, or do the spiral arms play a more active role?  Roberts
(1969) suggested that the spiral shock that occurs as the gas flows
through the potential minima triggers the star formation process in
spiral galaxies. Shock dissipation of excess kinetic energy can result
in the formation of bound structures which then collapse to form
stars. This of course can provide the dynamical mechanism we need to
form the observed molecular clouds. The formation of giant molecular
clouds in spiral arms can occur in one of two ways (e.g. Blitz, \&
Rosolowsky 2004): through the agglomeration of smaller molecular
cloudlets as they pass through the spiral arms (eg., Pringle \etal~2001)
or through the formation of molecules in dense atomic gas clouds
that themselves have been compressed in the spiral arms (Bergin \etal~2004, Elmegreen 1993).

In either formation scenario, we need to explain the observed
properties of GMCs.  GMCs are observed to contain highly supersonic
motions and a wealth of structure on all length scales (Larson~1981;
Blitz \& Williams 1999; Elmegreen \& Scalo~2004). The supersonic
motions are found such that the velocity dispersion $v$ varies with
length scale $R$ according to the relation $v\propto R^{1/2}$ (Larson
1981; Myers~1984; Miesch \& Bally 1993; Heyer \& Schloerb~ 1997; Heyer
\& Brunt~2004). A number of authors, on the basis on numerical
simulations, argue that it is these supersonic motions, maintained by
internal or external driving mechanism, which induce the observed
density inhomogeneities in the gas (Mac Low \& Klessen~2004; Elmegreen
\& Scalo~2004), and that it is therefore the supersonic motions which drive
star formation.  Suggested candidates for an internal driving mechanism include
feedback from low-mass star formation (Silk~1985) even though GMCs with and without
star formation have similar kinematic properties (Williams, Blitz \& McKee 2000).
External candidates include supernova and superbubbles (Wada \& Norman 2001; Elemegreen \& Scalo 2004). 
Although there is sufficient energy
in these events to explain the kinematics of the ISM and they can potentially generate the
correct velocity dispersion sizescale relation (Kornreich \& Scalo 2000), what is unclear is their relevance to star formation, and in particular to the triggering of star formation in spiral arms.

There are, as mentioned above, both observational and theoretical
reasons for supposing that molecular cloud lifetimes are quite short
(Elmegreen 2000; Pringle, Allen \& Lubow 2001; Hartmann,
Ballesteros-Paredes \& Bergin, 2001), being at most a few dynamical,
or kinematic (crossing) timescales.  If this is the case, then a
continuous source of turbulence is not required. Indeed, all that is
required is some initial impulsive input of energy. Moreover, in this
case, the motions do not need to be 
% turbulent it is evident that there is no need to regard the motions as
'turbulent' in the usual sense, as there is barely time for energy to
work its way through a cascade of eddies. What is the physical
mechanism which gives rise to such an impulse? Since star formation
occurs in spiral arms, the obvious place to look for such a mechanism
is in the spiral arms itself.

In view of all this, we suggest here an alternative origin for the
observed chaotic supersonic motions. It is well known that the
interstellar medium is highly structured on all scales (Lauroesch
\etal~2000; Dickey \& Lockman~1990; Elmegreen \& Scalo~2004). The
result of passing a shock through such clumpy gas is to give rise to
internal motions. The magnitude of the
motions so generated depends on the degree of inhomogeneity in the
pre-shock medium. We propose here that the passage of the clumpy
interstellar medium through a galactic spiral shock not only produces
the dense environment in which molecular clouds form (Cowie 1981; Elmegreen 1979, 1989, 1991),
but also gives
rise at the same time to their supersonic internal motions. We stress
again that these motions  are not really driven turbulence in the usual sense
of energy being passed from large eddies down through a spectrum of
smaller ones (Tennekes \& Lumley 1972). Rather, the passage of a shock through the clumpy
pre-shock medium leads to supersonic internal motions being generated at all
scales simultaneously. There have been several numerical studies of
the effects of spiral arms on gas flows. Some assume that the
pre-shock ISM is smooth and concentrate on the generation of
instabilities (e.g. Kim \& Ostriker~2002). Others consider global
effects of the galactic flows (Wada \& Koda~2004; Gittens \& Clarke
2004).  Here we present numerical simulations capable of following
both the galactic-scale gas dynamical response to the passage of a
spiral arm and the small-scale triggering of star formation, but
taking account of the inhomogeneity of the pre-shock ISM

\section{Calculations}

We use high resolution Smoothed Particle Hydrodynamics (SPH)
(Monaghan~1992, Benz \etal~1991) simulations to follow the passage of
interstellar gas through a spiral arm. The code has variable smoothing
lengths in time and in space and solves for the self-gravity of the
gas, when included, using a tree-code (Benz \etal~1991). 

\subsection{Preliminary simulations of clumpy shocks}

In order to demonstrate the basic idea, we first report on
calculations which compare uniform and clumpy gas clouds passing
through a shock induced by a one-dimensional sinusoidal potential.
These  demonstrate the role played by the clumpiness of the pre-shock
medium in generating the post-shock velocity dispersion. These
three-dimensional simulations follow the evolution of a non-self gravitating gas
through a one-dimensional potential of the form
\begin{equation}
\Phi = A \cos(k x),
\end{equation}
with $k=\pi/2$ and $A=80$ such that the potential has a minimum at
$x=2$.  Thus, the dimensionless velocity associated with the
potential, that is the velocity acquired by falling from the peak to
the trough, is $V_{\rm pot} = \sqrt{2 \times A} \approx 12.5 V$, where $V$ is the
code units for velocity. This potential was chosen to mimic the effect
of the passage through a spiral arm, but in a more controlled manner.
The cloud, represented by $2 \times 10^5$ SPH particles, initially has
length $3$ units, width $2$ units and height $2$
units. It starts off centred on the peak of the potential at
$x=0$. The internal sound speed of the gas is $ 0.3 V$, 
 and all particles are given an initial
velocity of $15 V$ in the $x$-direction, corresponding to a Mach
number of 50. The gas is assumed to remain isothermal throughout. 
These velocities are chosen so that the material shocks
itself as it climbs out of the trough of the potential in a manner
analgous to gas passing through the gravitational potential of a galactic
spiral arm (see below). We carry out three simulations, one in which the pre-shock
density is smooth, and two in which it is clumpy. The two clumpy runs
differ in the size of the clumps, with radii of $0.1$ and $0.2$
respectively. Both clumpy simulations are constructed of 1000 individual clumps
of 200 particles each. The clumps are bounded by an external pressure term.

\begin{figure}
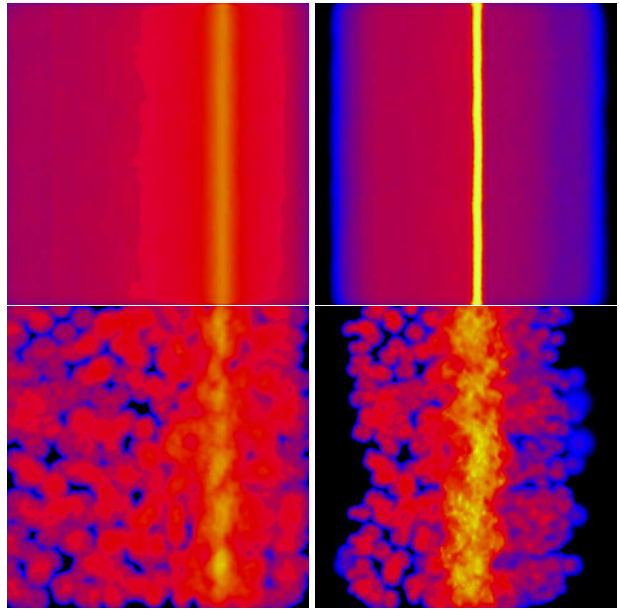

%\vspace{-0.5truein}
\centerline{\psfig{figure=CCrand030001-1.epsf,width=4.truecm,height=4.truecm}
\psfig{figure=CCrand050001-1.epsf,width=4.truecm,height=4.truecm}}
\centerline{\psfig{figure=CCRUN0030001-1.epsf,width=4.truecm,height=4.truecm}
\psfig{figure=CCRUN0050001-1.epsf,width=4.truecm,height=4.truecm}} 
\caption{\label{linshock} The column density distribution for the linear shock tests
are shown for the  uniform (top) and clumpy 
(bottom, with $r=0.1$ clumps) gas distributions. The left panels ($1.2 < x < 3.2,   -1 < y < 1$) show  the gas as it leaves the
potential minimum and begins to shock ($t=0.15$). The right panels ($2.6 < x < 4.6,     -1 < y < 1$)  show the gas distribution
when the shock is fully developed ($t=0.25$). The 
logarithmic column density levels range from  a minimum of 90 to a maximum of $9 \times 10^6$
in units of particles per unit area. }
\end{figure}

In Figure~\ref{linshock} we show the column density distributions of the smooth and small
clump simulations as they leave the potential minimum and start to shock ($t=0.15$)
and also at a later time ($t=0.25$) when the
shock is fully developed. The Mach number of the shock is $\approx 30$.
%\begin{figure*}
%\vspace{-0.5truein}
%\centerline{\hspace{-1.truein}
%\vbox{
%\centerline{\psfig{figure=CCrand030001.epsf,width=4.truecm,height=4.truecm}
%\psfig{figure=CCrand050001.epsf,width=4.truecm,height=4.truecm}}
%\centerline{\psfig{figure=CCRUN0030001.epsf,width=4.truecm,height=4.truecm}
%\psfig{figure=CCRUN0050001.epsf,width=4.truecm,height=4.truecm}} }
%\hspace{-2.truein}{\psfig{figure=cldshocksfn.ps,width=8.truecm,height=8.truecm} }\hfill}
%\caption{\label{linshock} }
%\end{figure*}
\begin{figure}
%\vspace{-0.5truein}
\centerline{\psfig{figure=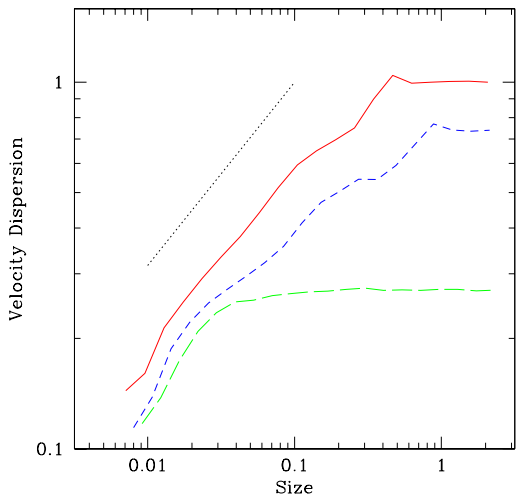,width=8.truecm,height=8.truecm}}
\caption{\label{cldshockvd}  The one-dimensional velocity dispersion of the post-shock gas 
as it passes through the linear potential is 
plotted against size-scale for a uniform (long-dashed) and two clumpy
shocks, with clumps of radii 0.1 (solid) and 0.2 (short dashed).
The velocity is in units where the sound speed is 0.3 and the
distance in units where the spacing between minima in the potential is 4 and the initial
configuration has a maximum spatial extent of 3. The dotted-line
illustrates the velocity dispersion size-scale ($v_{\rm disp} \propto R^{1/2}$) relation deduced for molecular clouds
(Larson~1981; Heyer \& Brunt 2004).}
\end{figure}
We compute the post-shock velocity dispersions from the velocity
component parallel to the pre-shock velocity. We average the velocity dispersion
over numerous distinct regions of a given size, all centred on SPH particles
with high density, and considering only particles with large densities indicating their location inside the shock.
 We then repeat this process on many different sizes
to determine how the induced velocity dispersion depends on
observed size-scale. We show these in Figure~\ref{cldshockvd} for the three
shock simulations at a time ($t=0.25$) when the shock is fully developed,
corresponding to the latter images in Figure~\ref{linshock}.
We see that the
uniform shock produces a constant, sub-sonic velocity dispersion on
all length-scales, except for those comparable to the individual particle separation.
 This is as expected, and demonstrates that we have
enough resolution to overcome the usual SPH $\sqrt{N}$ particle noise.
This velocity dispersion is related to the SPH handling of shocks
which leave a residual velocity dispersion as it is a particle based
method which is thus inherently clumpy on the smallest scales. 

In
contrast, and as expected, the clumpy shock gives rise to a
significant supersonic velocity dispersion in the post-shock gas. 
The post-shock velocity of a small element of gas depends on the amount
of mass it encounters, i.e.,  its mass loading.  Conservation of momentum and the
variation in the amount of mass loading due to the clumpy mass distribution 
thus lead to a velocity dispersion in the gas. 

Within a region of size-scale less than structures in the gas, each 
 individual parcel of gas encounters a similar column density in the shock.
 The uniform mass-loading then results in a zero or very small velocity 
 dispersion. As the size-scale of the region
 increases, the gas samples a variation in column densities due to 
 clumps. Then parcels of gas in the region will encounter differing 
 amounts of mass-loading
 as they enter the shock and thus differing decelerations.  The resulting
 velocity dispersion thus increases with the size of a region. 
 
 % A simple
 % model for the velocity dispersion based on mass loading in the above clumpy
 % shock adequately reproduces the gradient in the velocity dispersion.
 
This velocity dispersion increases with the size of a region as larger regions then 
encounter
a larger range of mass-loading in the shock.  In other words, larger regions
include more random samplings from the full range of mass loadings and thus 
post-shock velocities, resulting in a velocity dispersion which increases as $R^{1/2}$. 
The necessary condition being that the clumps have
a significant probability of collisions in the shock. 
Thus, from the smallest sizes  where the velocity dispersion is subsonic, the velocity dispersion increases
up to sizes of twice the clump diameter, and in so doing becomes highly supersonic.
 On larger scales, the velocity
dispersion is constant as the mass distribution is effectively uniform on these scales.
The 
simulation with smaller clumps (larger densities) produces a higher velocity dispersion and a
larger gradient than does the simulation with larger clumps, but it saturates
at a smaller size scale. The lower velocity dispersion in the larger clump
simulation is due
to the increased filling factor that reduces the dispersion
in mass loading that occurs in the shock.

% This velocity dispersion,
% due to the random nature of clump-clump collisions, increases with
% size-scale as these regions include increasing amounts of gas that
% encounter disparate amounts of mass in the shock. Conservation
% of momentum combined with the random amounts of mass-loading then
% produce a velocity dispersion that increases with size-scale. The 
% simulation with smaller clumps produces a higher velocity dispersion and a
% larger gradient than does the simulation with larger clumps. This is due
% to the increased filling factor of the larger clumps that reduces the dispersion
% in mass loading that occurs in the shock.

\subsection{Flow in a galactic potential}

The galactic potential we use is a combination of a two-armed spiral
potential taken from Cox \& Gomex (2002) of amplitude $n= 1$ atom
cm$^{-3}$, and a logarithmic potential (eg. Binney \& Tremaine 1987)
that provides a flat rotation curve with $v= 200$ km/s. The spiral
arms rotate as a fixed potential with pattern speed $2 \times 10^{-8}$
rad yr$^{-1}$.  In addition to the main scientific simulations, we
perform a number of control simulations in which self-gravity, or
alternatively the fluid shock, are turned off in order to assess the
physical importance of each in our results. The simulations were
carried out on the United Kingdom's Astrophysical Fluids Facility
(UKAFF), a 128 CPU SGI Origin 3000 supercomputer.

\subsubsection{Test particle simulation and the SPH initial conditions}

The initial positions of particles in the spiral hydrodynamic
simulations were taken from a test-particle simulation of orbits in
the galactic potential. The particles were initially placed on
tangential orbits with kinetic energies adapted from circular orbits
so as to be consistent with the spiral potential together with an
additional 5 \% dispersion. They were then allowed to evolve over
several orbits. The location of a significant density peak of size 100
pc in the spiral arm, at a galactic radius of $\approx 4$ kpc, was
chosen to identify $\approx 48000$ test particles of interest. The
earlier positions of these particles, before entering the spiral arm,
were then used to establish the initial conditions of the SPH
simulations. These test particles were given an additional 5 km/s
Gaussian vertical velocity dispersion and then each subdivided into 9
SPH particles to provide the required resolution of $4.3 \times 10^5$
SPH particles.  These initial conditions were used for our
simulations. There are three scientific simulations which we discuss
here, listed in Table~1. The first, Simulation A, uses the full initial conditions but
with particle masses chosen to provide a mean surface density of $\Sigma=1.0
$ \solmas\ pc$^{-2}$ and a total mass of $10^6$ \solmas. The second,
Simulation B, is identical except for  a lower surface density of $\Sigma = 0.1$
\solmas\ pc$^{-2}$ and total mass of $10^5$ \solmas.  The third,
Simulation C, uses only the central region of the above initial
conditions and then each particle is subdivided further, providing a
higher resolution simulation of $2.5 \times 10^6$ SPH particles and a
mass of $1.55 \times 10^5$ \solmas and  $\Sigma=1.0
$ \solmas\ pc$^{-2}$.  All subdivided particles had the
same kinematics as the original particles.

\begin{table} \caption{\label{Spiral shock simulations} }
\begin{tabular}{c|c|c|c}
Sim & $\Sigma (\solm {\rm pc}^{-2})$ & Mass (\solmas) & Npart   \\
\hline
\hline
A & 1 & $10^6$ &  $4.3 \times 10^5$   \\
B & 0.1 & $10^5$ &  $4.3 \times 10^5$  \\
C & 1 & $1.5 \times 10^5$ &  $2.5 \times 10^6$  \\
\hline
\end{tabular} \end{table}

\begin{figure}
%\vspace{-0.5truein}
\centerline{\psfig{figure=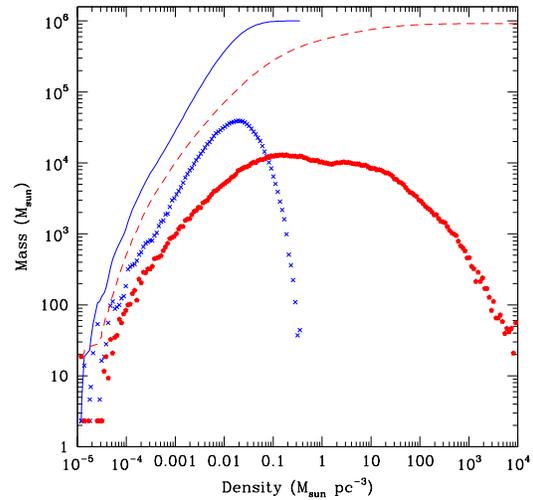,width=8.truecm,height=8.truecm}}
\caption{\label{denstruct} 
The distribution of mass as a function of gas density (points) is
plotted before passing through the spiral shock (crosses) and once
star formation has been triggered (filled pentagons) for Simulation
A. The corresponding cumulative distributions are shown as the solid
and dashed lines.}
\end{figure}

The gas is initially clumpy on scales of several pc with a mean
surface density of $1 \solm$ pc$^{-2}$, a volume-averaged density of
$2.5 \times 10^{-3} \solm $pc$^{-3}$, a median particle density of
$10^{-2}\solm $pc$^{-3}$ and peak densities in the clumps of $\approx
0.1 \solm$ pc$^{-3}$ (See Figure~\ref{denstruct}).  This clumpiness 
 corresponds roughly to the
properties of the interarm gas (Dickey \& Lockman 1990).  Self-gravity
is initially unimportant due to the large shear velocities. For
simplicity, the gas temperature was taken to be 100 K and remains
isothermal throughout the simulation. This corresponds to a sound
speed in molecular gas of $0.6$ km/s. We consider that the gas is
either in pre-existing molecular cloudlets (Pringle \etal~2001) or
that molecular formation is sufficiently rapid as to occur in the
shock compression (Bergin \etal~2004).  

%Simulations were performed
%with total gas mass of $10^5$ to $10^6$ \solmas\ of gas represented by
%$0.4$ to $2.5 \times 10^6$ SPH particles.

\begin{figure*}
%\vspace{-0.5truein}
\centerline{\psfig{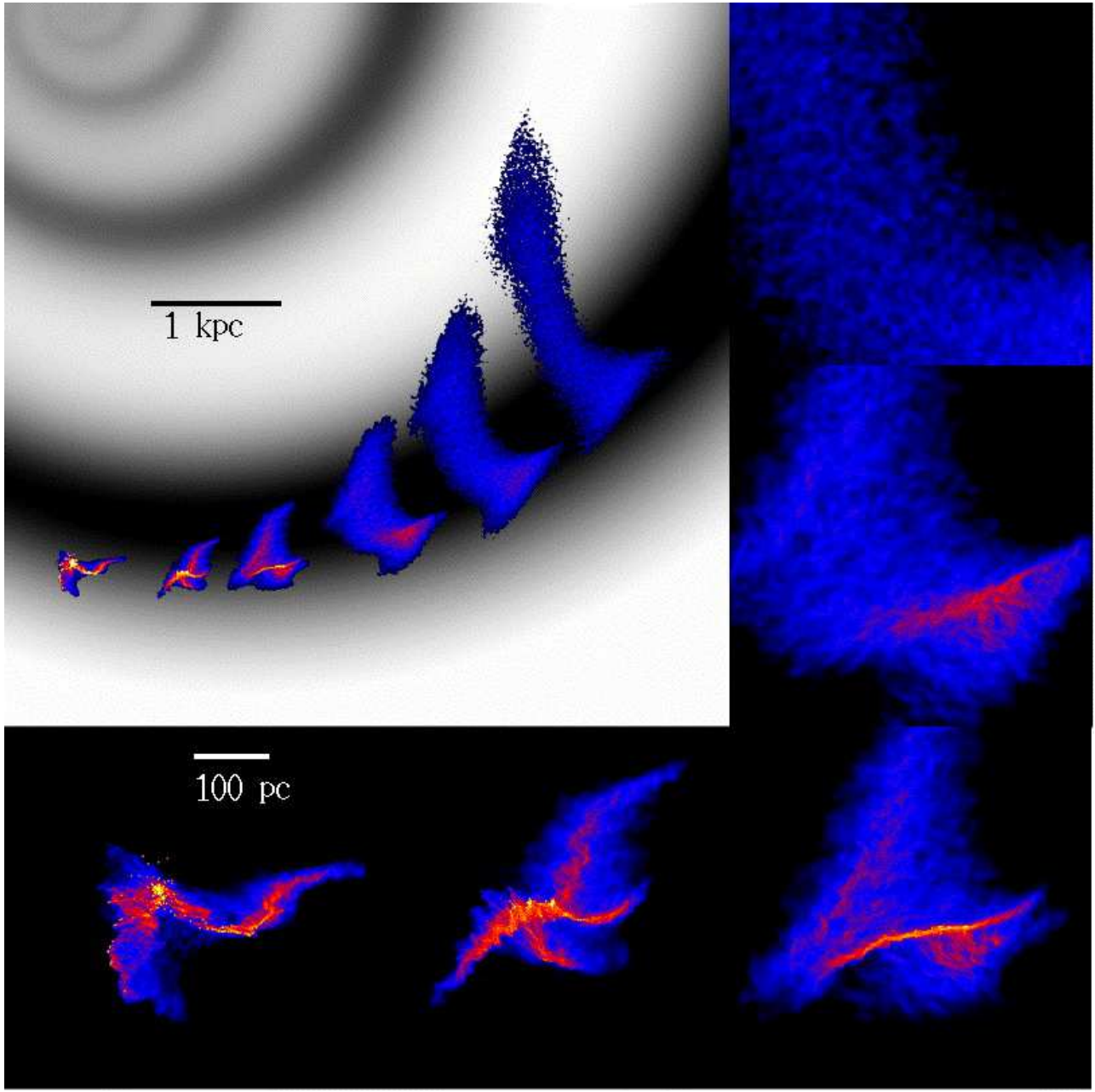}}
\caption{\label{spiraltrig} The evolution of cold interstellar gas ($m=10^6 \solm$) through a spiral arm is shown relative to the spiral potential of the
galaxy (upper left-panel) for Simulation A.. The minimum of the spiral
potential is shown as black and the overall galactic potential is not
shown for clarity.  The column densities of the gas vary from $5
\times 10^{-5}$ to $0.5$ g cm$^{-2}$. The gas is shown at times
$t=9\times 10^5$, $t = 9 \times 10^6$, $t =1.8 \times 10^7$, $ t = 2.3
\times 10^7$, $t= 2.8 \times 10^7$ and $t= 3.4 \times 10^7$ years from
the start of the simulation. The 5 additional panels, arranged
clockwise, show close-ups of the gas at the latter five times. The gas
is compressed in the shock and subregions become self-gravitating and
collapse to form groups of stars (from $t=2.3 \times 10^7$ years). The
cloud produces stars inefficiently as the gas is not globally bound. }
\end{figure*}

\subsubsection{Sink particles and star formation}

Star formation is modeled by the inclusion of sink-particles (Bate,
Bonnell \& Price 1995) that interact only through self-gravity and
through accretion of any gas particles that fall within their sink
radii. Sink-particle creation occurs when dense clumps ($\rho >
10^{-18} {\rm g cm}^{-3}$, $1.4 \times 10^ 3 \solm$ pc$^{-3}$) of
self-gravitating gas of size $\le 0.5$ parsecs are collapsing
(subvirial).  It should be noted that while the internal velocity
dispersion of these regions is probably underestimated as the
particles all lie inside one to two SPH kernels, the chosen
temperature of 100K mimics the internal support expected for lower
temperature gas with supersonic motions. This ensures that the star
formation modeled here is real.  The sink-particles have initial
(minimum) masses of order 25 to 100 \solmas\ and therefore cannot be
considered as individual stars but can be thought of instead as stellar
clusters. Accretion quickly increases the mass of these 'clusters' to
$10^2$ to $10^3$ \solmas) with maximum masses of $m \approx 10^4 $
\solmas. The detailed computation of star formation within these
clusters is beyond the scope of these investigations. 

\section{Triggering of star formation in the spiral shock}

\begin{figure*}
%\vspace{-0.5truein}
\centerline{\psfig{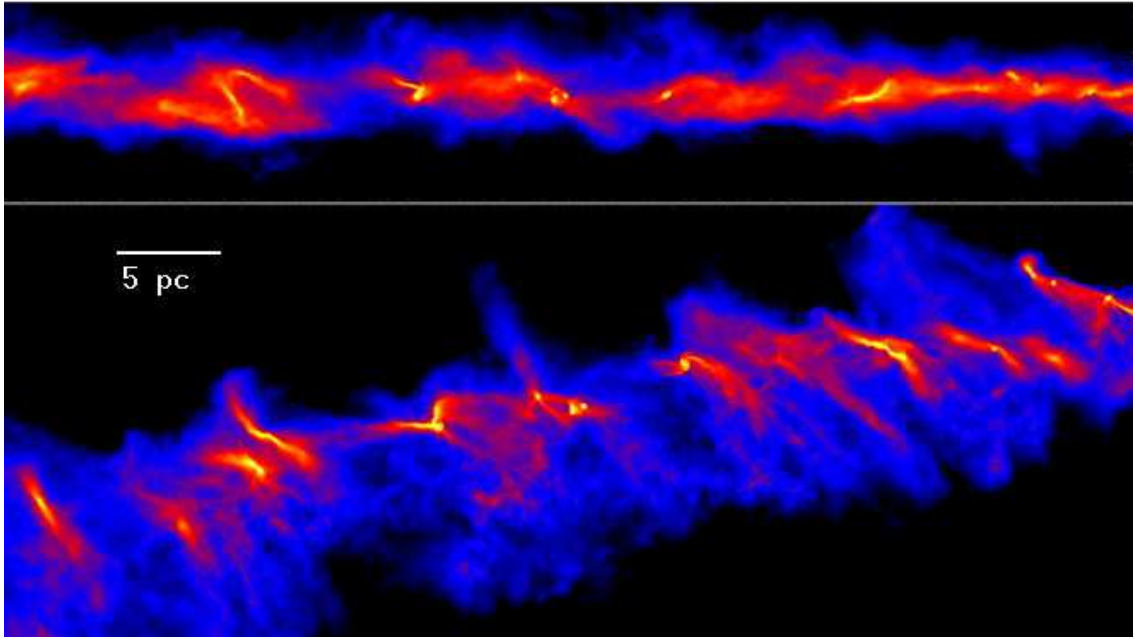}}
\caption{\label{GMCstruct} A 50 parsec region of the dense gas in the
spiralshock is shown in two projections, from within the disc of the
galaxy (upper panel ) and from above the disc (lower panel), taken
from a simulation of $3 \times 10^5 \solm$ modelled with $2.5 \times
10^6$ SPH particles.  The gas has fragmented into many
quasi-periodically-spaced dense regions with on-going star formation.
The fragmentation occurs due to the clumpiness of the pre-shock gas
and due to Rayleigh-Taylor like fluid instabilities in the
shock. Column densities of the gas vary between $5 \times10^{-4}$ and
1 g cm$^{-2}$.}
\end{figure*}

The evolution of the gas over 34 million years as it passes through
the spiral potential is shown in Figure~\ref{spiraltrig} for Simulation A with an
initial surface density of $\Sigma = 1 \solm$ pc$^{-2}$. The initially
clumpy, low density gas ($\rho \approx 0.01$ \solmas pc$^{-3}$,
$10^{-24}$ g cm$^{-3}$) is compressed by the spiral shock as it leaves
the minimum of the potential.  The shock forms some very dense
regions, which due to the accompanying dissipation of kinetic energy
and the increased importance of self-gravity, can become
gravitationally bound and thus collapse to form regions of star
formation. These regions, in excess of $10^3$ \solmas pc$^{-3}$ are
replaced by sink particles. Further accretion onto the sink particles
raises their masses to that of typical stellar clusters ($10^2$ to
$10^4$ \solmas).  Star formation occurs within $2 \times 10^6$ years
after sufficient densities are reached for the gas to be recognised as
a molecular cloud ($10^{-22}$ to $10^{-21}$ g cm$^{-3}$). The total
spiral arm passage lasts for $\approx 2 \times 10^7$ years.  The gas
remains globally unbound throughout the simulation and re-expands in
the post-shock region.
  
In our simulations, the total fraction of gas turned into stars varies
from between 5 and 30 \% depending on the simulation, total gas mass
and surface density. In reality this is likely to be an upper limit
since there are other mechanisms not included in our computations such
as the effects of winds and radiation from massive stars and supernova
explosions. We find that the total lifetime of the molecular clouds,
measured from the time that significant mass attains typical GMC
densities of $10^{-22}$ to $10^{-21}$ g cm$^{-3}$, to be of order
$10^7$ years, or a few dynamical times ($t_{dyn} \approx 4 \times
10^6$ years).

The shock dissipation in the spiral arm is essential to trigger the
star formation process.  We have also run simulations that exclude the
shock (with no SPH artificial viscosity) but include self-gravity; in
these we find that there is no induced star formation even with orbit
crowding and the increased importance of self-gravity in the spiral
arms.  Similarly, in Simulation B which is a full fluid simulation
with but with a lower initial surface density of $\Sigma=0.1 \solm$ pc$^{-2}$, 
only limited star formation occurs, producing  two 'star clusters'  before
the gas leaves the spiral arm and disperses. Triggering of star
formation thus requires both the dissipation of the excess kinetic
energy in the shearing flow and a critical threshold in surface
density in order for self-gravity to become important.

In addition to forming dense clouds in which star formation occurs,
the spiral shock forms structures which resemble the observed
structures in GMCs.  Detailed images of the GMCs formed in a spiral
shock are shown in Figure~\ref{GMCstruct} from Simulation C,  with the gas density
as in Simulation A, but at higher resolution simulation with $1.5 \times
10^5 \solm$ of gas, $\Sigma=1 \solm$ pc$^{-2}$, modelled with $2.5
\times 10^6$ SPH particles.  The projections of the dense gas along a
length of 50 pc, viewed from within and from above the disc of the
galaxy, are taken just after star formation has been initiated.  We
see significant amounts of substructure as the dense gas breaks up
into many components of several parsecs in size and separation.  The
dense regions have column densities typical to GMCs. This structure
can be understood as being due to the combination of the clumpy nature
of the pre-shock gas along with Rayleigh-Taylor like instabilities
which occur as gas continues to flow into the shock.

\begin{figure}
%\vspace{-0.5truein}
\centerline{\psfig{figure=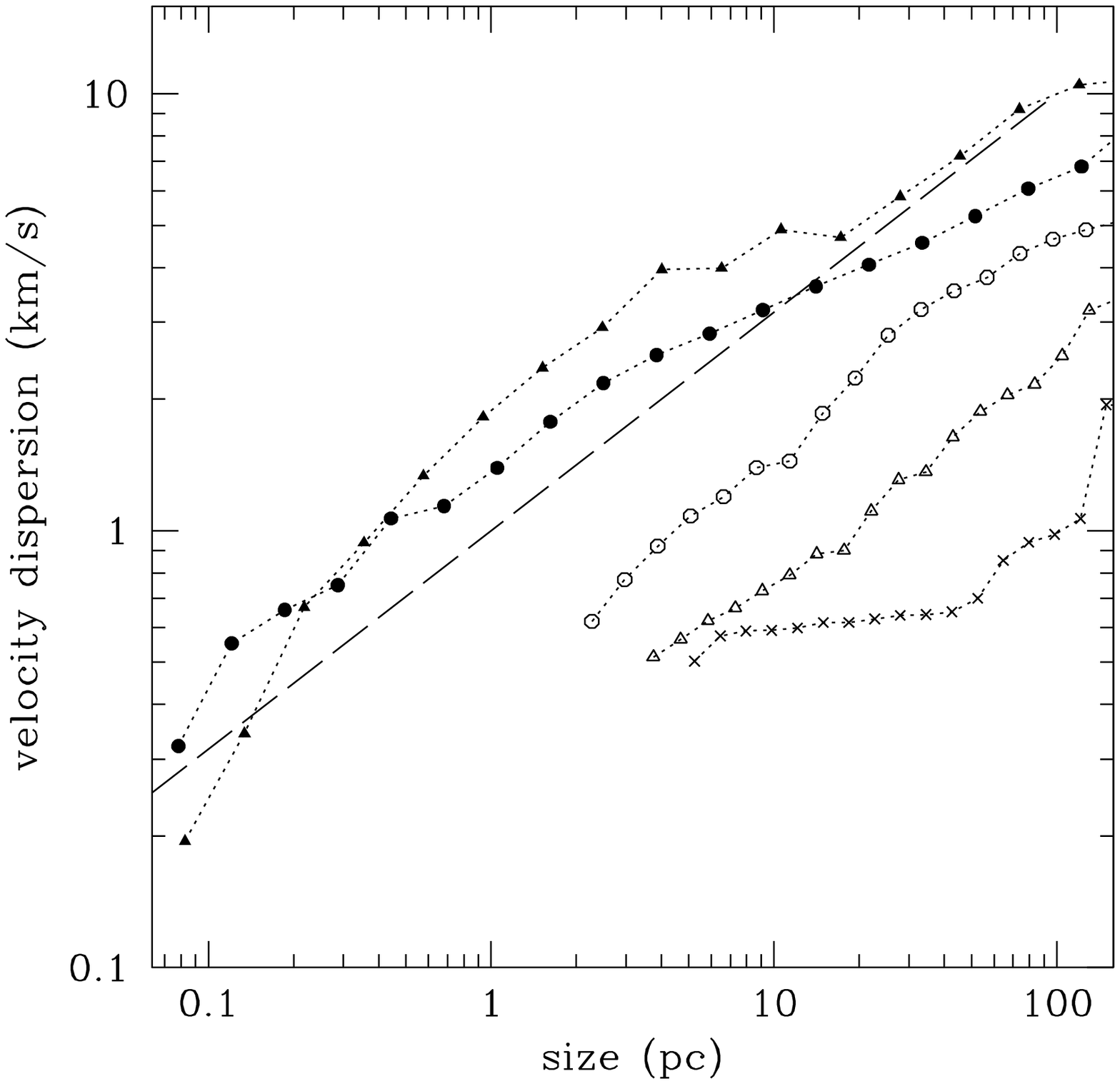,width=8.truecm,height=8.truecm}}
\caption{\label{larsonfig} The velocity dispersion is plotted as a
function of size at 5 different times during the passage of the gas
through a spiral shock (Simulation A).  The velocity dispersion is
plotted at $4.2 \times 10^6$ years (crosses), $1.4 \times 10^7$ years
(open triangles), $1.8 \times 10^7$ years (open octagans), $2.3 \times
10^7$ years (filled circles) and $2.7 \times 10^7$ years (filled
triangles) after the start of the simulation. Star formation is
initiated at $\approx 2.3 \times 10^7$ years. The velocity dispersion
is calculated from the velocities in the plane of the disc, which are
then averaged over many different regions of a given size that each
have central gas densities in excess of $ 10^{-21}$ g cm$^{-3}$. The dashed
line indicates the Larson relation for molecular clouds where $\sigma
\propto R^{-1/2}$ (Larson~1981; Heyer \& Brunt 2004).}
\end{figure}

\section{The generation of the internal velocity dispersion}

One of the most significant properties of GMCs is their supersonic
internal motions generally characterised by the internal velocity
dispersion (Larson 1981; Heyer \& Brunt 2004). Thus, in addition to
triggering star formation, any mechanism to explain the formation of
GMCs must be able to explain the origin of the internal velocity
dispersion and how it depends on the size of the cloud, or region
considered (Larson 1981; Heyer \& Brunt 2004). Here we discuss how
spiral triggering of star formation  produces these
kinematics when the pre-shock gas is clumpy.  The basic idea is that when structure exists
in the pre-shocked gas, the stopping point of a particular clump depends
on the density of gas with which it interacts.  thus some regions will
penetrate further into the shock, broadening it and leaving it with a
remnant velocity dispersion in the shock direction. The different elements of the gas pass
through shocks of different strengths and different orientations. The amount of mass-loading,
and hence deceleration that occurs in the shock is variable due to the clumps in the
pre-shock gas. Smaller scale regions  in the shock are likely to have more correlated
momentum injection as well as mass loading and thus small velocity dispersions. Larger
regions will have less correlation in both the momentum injection and mass loading 
such that there will be a larger dispersion in the post-shock velocity.
Furthermore, 
the spiral potential ensures that the gas follows converging streams into the
spiral shock such that there is a significant velocity component present
parallel to the shock front. Together, this imparts a
significant velocity dispersion in the post-shocked gas.

The evolution of the velocity dispersion in Simulation A as a function
of the size of the region considered is shown in Figure~\ref{larsonfig}. The
velocity dispersion is calculated from the velocity components in the
plane of the galaxy within a spherical (3-D) region. and then normalised
to produce a one-dimensional velocity dispersion, as would be observed.
 The velocity
dispersion in then averaged over multiple regions of the same size all
of which are centred on dense gas ($\rho \ge 10^{-21}$ g
cm$^{-3}$). We only use particles whose densities exceed a minimum value
of ($n \ge 10$ cm$^{-3}$; $\rho \ge 1.7 \times 10^{-23}$ g  cm$^{-3}$)
to ensure we only consider the gas that has entered the shock. 
Lower density gas, even if molecular, is unlikely to excite any
line emission.
The initially low velocity dispersion, of order the sound
speed $v_s \approx 0.6$ km/s, increases as the gas passes through the
spiral shock. At the same time, the velocity dispersion increases more
on larger scales than on smaller scales thus establishing a velocity
dispersion size-scale relationship which is approximately
\begin{equation}
v_{\rm disp} \approx  1  \left(\frac{R}{1 \rm pc}\right)^{1/2} {\rm km/s.}
\end{equation}  
By the time that star formation has been triggered, the gas contains
the characteristic kinematics of giant molecular clouds. The passage
of the clumpy gas through the shock produces both the power-law slope
and the magnitude of the velocity dispersion.

In the spiral shock, there is no distinction between the velocity dispersion inside a given
cloud and the {\sl intercloud} velocity dispersion. Thus, presumably the intercloud
velocity dispersion in the spiral shock will follow the same sizescale relation.
Once the clouds have passed through the shock they expand, with a simultaneous decrease
in  the internal velocity dispersion due to the dissipation of kinetic energy in the internal shocks. The velocity dispersion maintains a R$^{1/2}$ scaling but the magnitude
decreases. Such clouds, if observed, should somewhat decrease the slope of the
intercloud velocity dispersion sizescale relation. 
The caveat on this being that
such clouds are likely to be increasing difficult to detect in this epoch due to  their
decreasing  surface densities.

This driving of the internal velocity dispersion due to the spiral
shock occurs even for simulations where self-gravity is not included.
We can thus exclude that the velocity dispersion in our simulations is simply a
reflection of the virialised nature of self-gravitating clouds. These
clouds can be far from virialised, in fact completely unbound, and
still display the same velocity dispersion.  Thus  we can
fully attribute the generated velocity dispersion, and its dependency on the
length scale considered, to the passage of the clumpy gas through the
spiral shock.

\section{Discussion}

The scenario presented here, of a dynamical triggering of star
formation by the passage of inhomogeneous interstellar gas through a
spiral shock, is consistent with the emergent viewpoint that star
formation is a relatively fast process whereby clouds form out of the
ISM, produce stars and then disperse, all on their local dynamical, or
crossing, times (Elmegreen~2000; Hartmann \etal~2001). Few GMCs are
observed without ongoing star formation suggesting that star formation
must occur soon after the clouds form. Estimates based on the ages and
distribution of young stellar populations imply that star formation
itself occurs on a dynamical timescale (Elmegreen~2000) while the lack
of molecular gas around older systems suggest that the clouds disperse
after at most a few dynamical timescales (Leisawitz \etal~1989). From
our simulations we estimate lifetimes of the dense clouds of order
$10^7$ years, or a few dynamical times ($t_{dyn} \approx 4 \times
10^6$ years), with pre-star formation lifetimes of order $2\times
10^6$ years.  We find star formation efficiencies in the range 5--30
percent, but note that these are likely to be reduced by including the
effects of feedback from young stars and/or magnetic fields.

Our simulations also demonstrate that the observed internal kinematics
of GMCs can be caused simply by the passage of an initially clumpy ISM
through the spiral shock.  There is then no need for any  internal or external driving mechanism 
for the 'turbulence'
in terms of magnetic instabilities, gravitational instabilities,
stellar outflows, winds or supernovae (Mac Low \&
Klessen, 2004). The velocity dispersion is
generated at all scales simultaneously (Brunt~2003) and thus no turbulent cascade 
of energy from larger to smaller lengthscales is required.  Once generated, these
internal motions are likely to evolve as in  simulations of decaying supersonic 
turbulence  (Mac Low \etal 1998). 
A caveat to this notion of decaying turbulence is that only 25 per cent of the
gas in the dense clouds has centre of mass motions  with radii of curvature
of 10 pc or less.  Thus the majority of the internal motions in the dense clouds 
do not show evidence for  significant eddy motions on sizescales of molecular clouds.

Furthermore, we have seen that GMCs
do not need be globally gravitationally bound in order to explain
their properties or their evolution.  This relaxation of a generally
held assumption (that the GMCs are in virial equilibrium) permits a
straightforward explanation for the low star formation efficiency in
GMCs. The vast majority of the gas is never gravitationally bound and
thus does not undergo star formation.  Local regions in unbound clouds
can dissipate sufficient kinetic energy to become gravitationally
bound and thus collapse to form stars (Clark \& Bonnell~2004). These
regions then provide the initial conditions for detailed studies of
the star formation process (Bate, Bonnell \& Bromm 2003; Bonnell, Bate
\& Vine 2003). A recent simulation of an unbound GMC has shown that
star formation efficiencies on the order of 10 per cent are a natural
outcome of the unbound nature of the clouds (Clark \etal~2005).

Although our models are explicitly made with two-armed grand-design spiral features, the
spiral shock origin for star formation could equally apply to flocculent spiral galaxies. In such systems
any gravitational instabilities will be sheared into local spiral arms (Elmegreen, Elmegreen \& Leitner 2003) as occurs in any weakly self-gravitating disc (Rice \etal 2003). Shocks due to the passage of gas through these spiral arms
can then trigger star formation much in the way described in this paper.

\subsection{Comparison with observations}

Detailed comparisons with observations of spiral galaxies (Tilanus \& Allen 1993; 
Allen \etal~1997; Smith \etal~2000; Schinnerer \etal~2003) will require global disc
calculations (Dobbs \etal, in preparation) and should ideally include more physical processes such as a
realistic equation of state, feedback from young stars, and the
effects of magnetic fields. 
Notwithstanding, there are several 
predictions from this scenario which can be tested
observationally.

First, the mechanisms detailed here is of a clumpy gas that forms GMCs
due to the passage through a spiral shock. The implication is that GMCs should
reside primarily in spiral arms of galaxies. This is consistent with surveys
of the outer Galaxy (Heyer \etal~1998) which finds an arm-interarm density ratio of GMCs in the Perseus arm to be 28:1 (Heyer \& Terebey 1998). The overall distribution of GMCs in the outer galaxy also
reinforces the impression that they are confined to spiral arms (Heyer, personal communication)\footnote{The presence of GMCs in spiral arms is, at least in our Galaxy, not surprising as the spiral arms are generally defined
by the presence of star formation and hence of star forming GMCs.}.  The global distribution of CO gas in the Galaxy is also best explained
as being predominantly in spiral arms (Bissantz, Englmaier \& Gerhard 2003).

 Second, the internal velocity dispersion is driven
primarily at the shock itself. Therefore the gas should have
relatively low internal velocity dispersions prior to the shock that
increase rapidly at the shock front and decay monotonically
thereafter.  

Third, the strength and scaling relations of the
velocity dispersion should be independent of the proximity of young
stars or other potential sources of turbulence. This appears to be the
case in the Carina giant molecular cloud (Zhang \etal~2001).  

Fourth,
the shock-driven origin of the GMCs implies that they should have
sharper edges on the upwind side where fresh material is flowing into
the shock, while the downwind edge to the GMCs is smoother due to the
gas dispersal. This can be seen in Figure~\ref{GMCstruct}. 

 Fifth, a potential
implication of this work is that at least some component of the the
velocity dispersion seen in the inter-arm gas (Dickey \& Lockman~1990)
is due to this pumping process in the spiral arms. The structure and
kinematics that result from the spiral shock should then feed directly
into the next spiral passage.

\section{Conclusions}

The triggering of star formation by the passage of clumpy gas through
a spiral arm can explain  many of the observed properties of
star forming regions. The clumpy shock reproduces the
observed kinematics of GMCs, the so-called 'Larson' relation. There is
no need for any internal driving of the quasi-turbulent random
motions.  The shock forms dense structures in the gas which become
locally bound and collapse to form stars. The clouds are globally
unbound and thus disperse on timescales of $10^7$ years, resulting in
relatively low star formation efficiencies.

\section*{Acknowledgments}
The computations reported here were performed using the U.K. Astrophysical
Fluids Facility (UKAFF). We thank Bruce Elemegreen, Mark Heyer, Chris Brunt and the
referee John Scalo for helpful comments.

\end{document}